\documentclass[pdflatex,sn-nature]{sn-jnl}

\usepackage{amsmath}
\usepackage{amssymb}
\usepackage{amsfonts}
\usepackage{bbm}
\usepackage{graphicx}
\usepackage{multirow}
\usepackage{booktabs}

\providecommand{\ket}[1]{\left|#1\right\rangle}
\providecommand{\bra}[1]{\left\langle#1\right|}

\begin{document}

\title[Metrological steering advantage via stochastic filtering]{Experimental activation of hidden metrological steerability in quantum channels via stochastic filtering}

\author*[1]{\fnm{H. S.} \sur{Karthik}}\email{karthik.hs@ug.edu.pl}
\author[2,3]{\fnm{J.} \sur{Berm\'udez}}
\author[2,4]{\fnm{D.} \sur{Montoya}}
\author[2,3]{\fnm{J.} \sur{Parra-Past\'en}}
\author[2]{\fnm{F. M.} \sur{Quinteros}}
\author[1]{\fnm{Akshata} \sur{Shenoy H.}}
\author[1]{\fnm{M.} \sur{Paw{\l}owski}}
\author[2,3]{\fnm{S. P.} \sur{Walborn}}
\author[2,3]{\fnm{G.} \sur{Lima}}
\author*[2,3]{\fnm{E. S.} \sur{G\'omez}}\email{estesepulveda@udec.cl}

\affil[1]{\orgdiv{International Centre for Theory of Quantum Technologies},
  \orgname{University of Gda\'nsk},
  \orgaddress{\street{Prof. Marii Janion 4}, \postcode{80-309}, \city{Gda\'nsk}, \country{Poland}}}

\affil[2]{\orgdiv{Departamento de F\'isica},
  \orgname{Universidad de Concepci\'on},
  \orgaddress{\street{Casilla 160-C}, \city{Concepci\'on}, \country{Chile}}}

\affil[3]{\orgdiv{Millennium Institute for Research in Optics},
  \orgname{Universidad de Concepci\'on},
  \orgaddress{\street{Casilla 160-C}, \city{Concepci\'on}, \country{Chile}}}

\affil[4]{\orgdiv{Departamento de Ingenier\'ia Inform\'atica y Ciencias de la Computaci\'on},
  \orgname{Universidad de Concepci\'on},
  \orgaddress{\postcode{4070409}, \city{Concepci\'on}, \country{Chile}}}

\abstract{Quantum steering-assisted metrology allows a trusted party (Bob) to improve local phase estimation by exploiting classical information communicated by an untrusted party (Alice). In realistic scenarios, however, the channel is subject to noise that degrades steerability and suppresses the associated metrological advantage. Here, we demonstrate that stochastic pre- and post-filtering operations can revive the violation of the metrological steering inequality (MSI) in phase-covariant amplitude-damping channels where it is initially absent, thereby activating a hidden metrological steerability and restoring their usefulness for quantum metrology. Specifically, the critical damping parameter beyond which the MSI violation is suppressed is raised from $\gamma_c \approx 0.5$ to $\gamma_c \approx 0.75$ for the largest filter strength considered. We verify this in a photonic experiment, where polarization qubits carrying the encoded phase are transmitted through an amplitude-damping channel implemented with a two-path Sagnac interferometer, and the stochastic operations are realized with glass plates at the Brewster angle. By tuning the filter strength and scanning the damping parameter over its full range, we observe the revival of the MSI violation, obtained at the cost of a reduced success probability inherent to the stochastic nature of the filtering operations. Our results establish that metrological utility and steerability can be activated through stochastic operations even when these properties are absent in the original noisy channel, extending filtering-based activation from witnesses of nonclassicality to a certified and operationally quantified metrological advantage.}


\maketitle

\section{Introduction}
\noindent Quantum steering occupies an intermediate position in the hierarchy of quantum correlations, lying between entanglement and Bell nonlocality~\cite{wise07}. Originally introduced by Schrödinger \cite{Schron36} in the context of the Einstein–Podolsky–Rosen paradox \cite{eins35}, steering refers to the ability of one party (Alice) to \textit{remotely} prepare,
or \textit{steer} the set of conditional states of another party (Bob) through her local measurements. Steering is formulated operationally in terms of assemblages and the impossibility of describing the observed correlations using a local hidden state (LHS) model \cite{wise07, roope20}. Beyond its foundational significance, steering has emerged as a useful resource in one-sided device-independent protocols, where one party’s devices are trusted (completely characterized) while the other’s are not \cite{branciard12,chen21}. Recent works have studied steering in a variety of settings, including temporal steering \cite{chen2014temporal,karthik2015joint,bartkiewicz2016experimental, PhysRevA.97.032301}, channel steering \cite{piani2015channel}, continuous variable \cite{Kimble1998,walborn11,Olsen2013,He2015}, high-dimensional \cite{gomez21} and multipartite steering \cite{he2012genuine,arms2015,cavalcanti2015,mattar17,taddei21b}, loophole-free \cite{bennet12, wittmann2012loophole} and robustness \cite{sabine20} of steering criteria, distillation \cite{nery2020} and revival of steering under noisy conditions~\cite{Xiaowei21}, one-sided device independence in quantum cryptography~\cite{branciard12,chen21} and measurement-device-independent scenarios~\cite{kocsis2015mdisteering, zhao20}. These works highlight that the operational content of steering lies in the structure of the conditional states rather than strictly in spatial nonlocal correlations.

A particularly important development in the study of steering was the formulation of Reid’s criterion, which showed that steering can be witnessed through the violation of inferred uncertainty bounds, thereby providing an operational way to demonstrate the EPR paradox using continuous-variable systems \cite{reid1989demonstration,reid09,he2012genuine}. Reid’s criterion connected steering to conditional uncertainty relations. Since parameter estimation precision is fundamentally limited by uncertainty relations and quantified by Fisher information, this connection reveals a deeper link: the quantum Fisher information optimized over a steering assemblage (the set of conditional states remotely prepared by Alice's incompatible measurements) can exceed the bound imposed by any local hidden state model, establishing steering as a genuine resource for enhanced parameter estimation. This insight gave rise to the framework of steering-assisted quantum metrology \cite{yadin2021metrological}.



In steering-assisted quantum metrology \cite{yadin2021metrological,gianani2022witnessing,lee2023}, a trusted party (Bob) aims to estimate a parameter encoded locally on his system while conditioning his measurement strategy on classical information obtained from an untrusted party (Alice). The main idea is that the assemblage generated by Alice’s measurements can improve the precision of phase estimation beyond what is achievable with unsteerable states. This advantage can be quantified using the quantum Fisher information associated with the conditional states, leading to metrological steering inequalities that bound the achievable precision under local hidden state models. Violating these inequalities certifies that the observed metrological advantage arises from steering. Importantly, the metrological task depends only on the assemblage available at Bob, which allows the framework to be implemented not only in spatially separated systems but also in temporal scenarios described by prepare-and-measure schemes. 

In realistic scenarios, however, the steering-assisted quantum metrology process is inevitably affected by environmental noise, which degrades both steerability and metrological precision. In many practical metrological platforms, such as optical interferometry, atomic sensing, and solid-state qubits, decoherence and loss act on the probe during the phase estimation process, such that the sample is better described as a noisy channel. Thus, from both a fundamental and practical perspective, we therefore ask whether metrological usefulness can be recovered in such noisy scenarios. Recent studies have shown that quantum correlations such as Bell nonlocality \cite{popescu1995bell, gisin1996hidden, flavien2013genuinehidden, Marcotulio2015inequivalence} and steering \cite{PhysRevA.99.030101, roope20, nery2020, taddei21b,huan22, yu22steering, PhysRevA.109.022411, hsieh2023characterisation, ducuara2022hidden} can be activated or revealed through local filtering or stochastic operations, even when they are not directly observable in the original state or channel. In dissipative two-qubit systems, for instance, hidden nonlocality, steering, and usefulness for teleportation can all be extracted by local filtering, and the strength of the recovered correlations trades off against the probability that the filtering succeeds \cite{ducuara2022hidden}. These results raise a natural question: whether noisy and initially unsteerable channels can exhibit hidden steerability under suitable stochastic operations \cite{kim2012protectingentanglement, ku2021hiddennonmacrorealism, karthik2025nrqrac}, and whether such activation can restore their usefulness for phase estimation. Addressing this question is particularly relevant for experimental platforms, where noise is unavoidable and stochastic processing is often feasible. 

In particular, we note that weak measurement and measurement reversal techniques have been used for noise mitigation against the amplitude-damping channel for over a decade. In Ref.~\cite{kim2012protectingentanglement}, the authors have demonstrated decoherence suppression via weak measurement and reversal, and in Ref. \cite{PhysRevA.95.042342}, similar ideas were applied to entanglement protection under amplitude damping and generalized amplitude damping channels. More recently, Ref.~\cite{ku2021hiddennonmacrorealism} considered reviving the quantumness of the amplitude damping channel, as witnessed by violation of a macrorealism inequality, and Ref.~\cite{karthik2025nrqrac} reported the revival of a dimension-witness violation using pre- and post-stochastic operations when amplitude damping was acting on the channel.  Additional relevant work on entanglement distillation using local filtering techniques includes Ref.~\cite{Kwiat2001experimental,PhysRevLett.96.220505,PhysRevLett.100.090403,Gomez2018_tunable}. 

Our approach adopts a similar philosophy, but differs from the works above in the object being recovered. Those results certify that a state or a channel retains some form of nonclassicality, whether nonlocality, steerability, or macrorealism violation, without attaching that certification to a task the experimenter is trying to perform. The MSI is different: its violation certifies a quantifiable gain in phase-estimation precision over any local-hidden-state strategy. What is recovered here is therefore not a signature of quantumness but an operationally defined metrological resource. We demonstrate that stochastic operations can unveil hidden metrological steerability in noisy phase-encoded channels, restoring their utility for quantum metrology through renewed violation of the metrological steering inequality (MSI). Our work establishes that this restored capacity represents a certified metrological steering advantage: Alice's ability to remotely prepare Bob for enhanced phase estimation, which is distinct from and not guaranteed by the mere recovery of entanglement or fidelity. These results pave the way for noise-adapted metrological protocols in one-sided device-independent settings, where stochastic processing provides a practical route to certifying quantum advantage without requiring entanglement recovery.


\section{Results}

\subsection{Quantum steering}
Quantum steering is conventionally framed as an information-theoretic task: an untrusted party, Alice, holds a source producing pairs of photons, retaining one and sending the other to a trusted, spatially separated Bob. Alice’s objective is to convince a skeptical Bob that they share an entangled state. To verify this, Bob asks Alice to perform measurements on her subsystem and checks whether her measurement outcomes determine the conditional states of his subsystem in a way that cannot be explained by a local hidden-state (LHS) model. Let Alice’s possible measurement settings be denoted by
$x \in \mathcal{X} = \{1, \ldots, M_A\}$, with corresponding \textit{positive operator valued measures }(POVMs) elements
$\mathrm \{E_{a|x}\}_{a=1}^{m_A}$, and the outcomes are labeled by
$a \in A = \{1, \ldots, m_A\}$. Bob measures his subsystem with a POVM $M = \{M_b\}$, with outcomes labeled by $b$. Conditioned on measuring $x$ and obtaining an outcome $a$, Bob obtains a set of unnormalized quantum states known as assemblages $\{\sigma^{B}_{a|x}\}_{a,x}$, where $\sigma^{B}_{a|x}=\mathrm Tr_{A}[(E_{a|x} \otimes \mathbbm{1}) \rho_{AB}]$; $\rho_{AB}$ represents the joint state of Alice and Bob. If the  assemblages at Bob's end admit a decomposition known as LHS description
\begin{equation}
    \sigma^{B}_{a|x}=\sum_\lambda p(\lambda)P(a|x,\lambda) \rho^{B}_{\lambda},
    \label{eq:LHS}
\end{equation}
where $0\le p(\lambda)\le 1$, $\sum_{\lambda} p(\lambda)=1$,  $0\le P(a|x,\lambda)\le 1$ and $\sum_{a}P(a|x,\lambda)=1$, then the assemblage admits a LHS model and no steering is demonstrated. In other words, the LHS model provides a classical description of the assemblages.
Furthermore, the assemblages are non-signaling, that is
\begin{equation}
    \sum_{a} \sigma^{B}_{a|x}= \sum_{a} \sigma^{B}_{a|x'} = \rho^B, \quad \forall x,x',
\end{equation}  
and $(\text{Tr}\sum_{a} \sigma^{B}_{a|x}) = 1$ \,\,$\forall x$; $\rho^B=\text{Tr}_A(\rho_{AB})$. If Bob's assemblages do not admit a LHS description, then steering is witnessed. Importantly, quantum steering requires that Alice's measurements be incompatible~\cite{quintino14,roope14}.

Beyond the spatial case, the temporal counterpart of quantum steering~\cite{chen2014temporal, karthik2015joint} proceeds as follows. Initially, Alice prepares the state $\rho$. Alice then measures $x$ on $\rho \rightarrow \rho_A$ at an instant of time $t_A$ with outcomes $a$. The measurements corresponding to the inputs $x$ can be a set of incompatible POVMs $\{E_{a|x}\}$ such that Bob receives the post-measurement assemblage
$\{\sigma^{B}_{a|x}\}_{a,x}$. Each unnormalized state in the assemblage is given by
\begin{equation}
    \sigma^{B}_{a|x} = \sqrt{E_{a|x}}\rho_{A}\sqrt{E_{a|x}}
\end{equation} where $\sum_a E_{a|x} = \mathbbm{1}$ and $E_{a|x} \ge 0$. To verify that Alice has transmitted a steering assemblage, Bob measures $M$ and obtains outcomes $b$ at a later time $t_B (t_B > t_A)$. Temporal quantum steering is observed from Alice to Bob if Alice performs incompatible measurements \cite{karthik2015joint} and the resulting assemblage cannot be decomposed in the LHS form. The unnormalized states $\sigma^B_{a|x}$ play the same role as in the spatial case, and the corresponding normalized conditional states are $\rho^B_{a|x} = \sigma^B_{a|x}/P(a|x)$, with $P(a|x) = \mathrm{Tr}[\sigma^B_{a|x}]$.

We now exploit the assemblage structure developed above as a resource for parameter estimation. Crucially, since the metrological task depends only on the assemblage available at Bob and not on whether the correlations are spatial or temporal, the framework we develop in the next section applies equally to the spatial scenario described here and to the temporal prepare-and-measure scheme realized in our photonic implementation.

The temporal scenario retains the one-sided structure of the spatial one. Two distinct devices are involved: an untrusted preparation device, which on input $x$ emits a system in an uncharacterized state, and a trusted measurement device operated by Bob. Bob never assumes a description of the preparation device, and the LHS decomposition of Eq.~(\ref{eq:LHS}) is precisely the statement that this device behaves as a classical source of pre-existing states. One difference from the spatial case should be made explicit. There, spacelike separation enforces the absence of communication between the parties as a matter of physics, whereas in the prepare-and-measure setting the preparation device physically transmits the system to Bob. The one-sided device-independent reading therefore carries the additional assumption, standard in temporal steering and in prepare-and-measure protocols, that the transmitted system carries no side information beyond the qubit degree of freedom under consideration.

\begin{figure}[th!]
\centering
\includegraphics[width=0.8\textwidth]{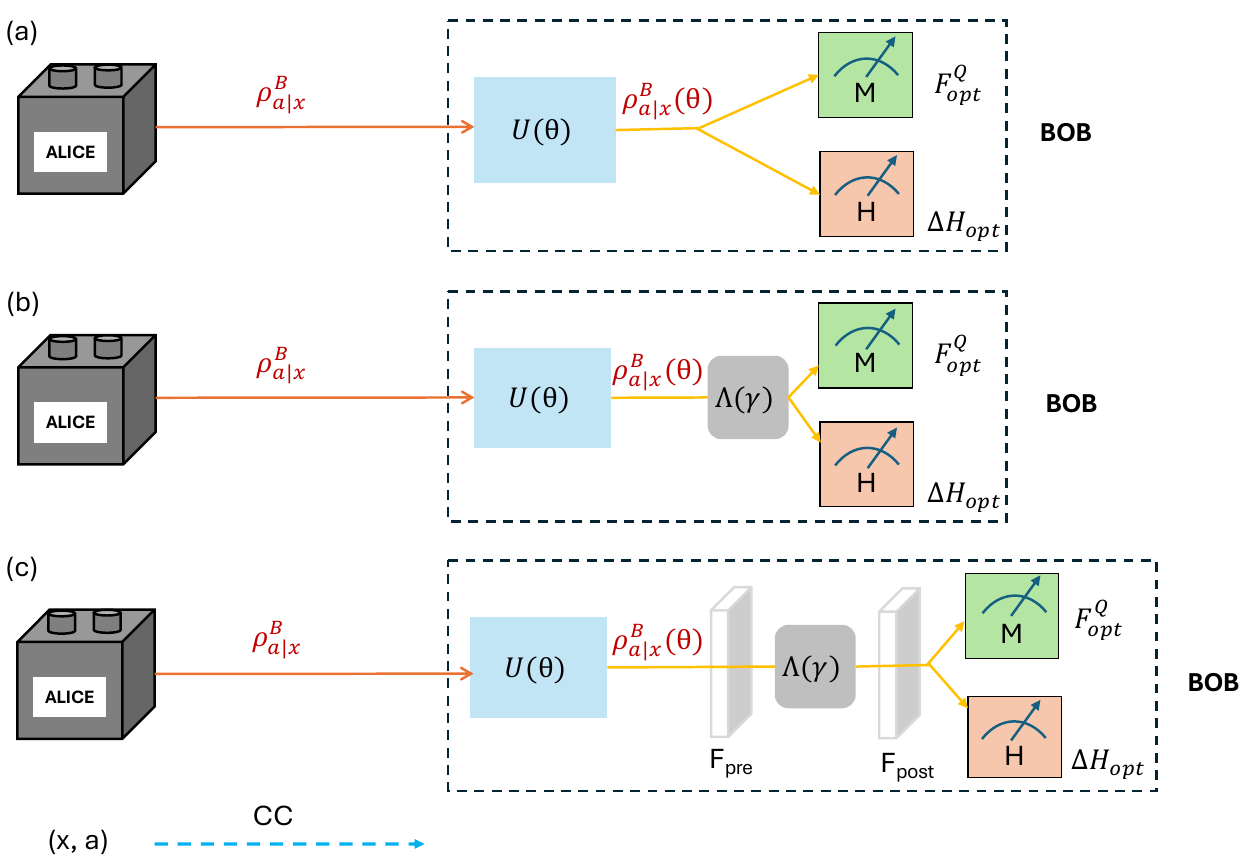}
\caption{Schematic of the quantum steering-assisted metrology protocols considered in this work. In (a), the ideal scenario. Alice performs different measurements and classically communicates the measurement setting and outcome $(x, a)$ to Bob, thereby remotely preparing the corresponding assemblages $\{\rho^B_{a|x}\}$. Bob then encodes a parameter $\theta$ through a unitary phase evolution generated by a Hamiltonian $H$. He either performs a Fisher information-optimizing measurement $M$ for estimating $\theta$ or measures $H$ directly and evaluates the variance. In (b), noisy version of the above protocol where Bob encodes the parameter in the same manner, but the phase encoded state undergoes a noisy quantum channel $\Lambda_{\theta}$ before the final measurements. (c) shows a noise-robust steering-assisted metrology protocol: Bob applies the stochastic operation $F_{pre}$ to the states $\rho^B_{a|x}$ before transmission across the noisy quantum channel $\Lambda_{\theta}$. The noisy state received by Bob undergoes $F_{post}$ before the final measurement for testing the MSI.}
\label{fig:mainfig}
\end{figure}


\subsection{Quantum steering for metrology}
We now consider quantum steering in a metrological scenario~\cite{lee2023}.  Suppose that, under ideal conditions, Bob applies a unitary operation $U(\theta) = e^{-iH\theta}$, which encodes an unknown phase parameter $\theta$ 
onto his conditional state $\rho^{B}_{a|x}$, conditioned on Alice’s 
measurement outcome $a$ for setting $x$, communicated classically. 
The encoded state at Bob’s side is given by
\begin{equation}
    \rho^{B}_{a|x}(\theta)
    = U(\theta)~ \rho^{B}_{a|x}~ U^{\dagger}(\theta)
    = e^{-iH\theta}\, \rho^{B}_{a|x}\, e^{iH\theta},
\end{equation}
where $H$ is the Hamiltonian generating the phase evolution. The corresponding assemblage at Bob’s side is
\begin{equation*}\left\{\sigma^{B}_{a|x}(\theta)
    =  P(a|x)\,\rho^{B}_{a|x}(\theta) \right\}_{x,a,\theta}.\end{equation*}

Bob may perform a measurement $M=\{M_b\}$ to estimate $\theta$, 
or equivalently probe the generator $H$. The ultimate precision is 
quantified by the quantum Fisher information (QFI) \cite{giovannetti2006,demkowicz2014, yadin2021metrological}, 
defined as
\begin{equation}
    \mathbb{F}_Q^{\mathrm{opt}}
    := \max_{x} \sum_{a} P(a|x)\, \mathbb{F}_Q\!\left(\rho^{B}_{a|x}(\theta)\right),
\end{equation}
where the optimization is over Alice’s input setting $x$. The QFI is 
\begin{equation}
    \mathbb{F}_Q\!\left(\rho^{B}_{a|x}(\theta)\right)
    := \mathrm{Tr}\!\left[\rho^{B}_{a|x}(\theta)\, L_{\theta}^{2}\right],
\end{equation}
with the symmetric logarithmic derivative (SLD) $L_{\theta}$ defined implicitly by
\begin{equation}
    \partial_{\theta} \rho^{B}_{a|x}(\theta)
    = \frac{1}{2}\left(L_{\theta}\rho^{B}_{a|x}(\theta)
    + \rho^{B}_{a|x}(\theta)L_{\theta}\right).
\end{equation}

For a fixed measurement $M=\{M_b\}$, the outcome probabilities follow Born’s rule,
\begin{equation}
    p(b|M,\rho^{B}_{a|x}(\theta))
    = \mathrm{Tr}\!\left[M_b\, \rho^{B}_{a|x}(\theta)\right].
\end{equation}

The quantum Fisher information upper-bounds the classical Fisher information \cite{braunstein1994statistical},
\begin{equation}
    \mathbb{F}_Q\!\left(\rho^{B}_{a|x}(\theta)\right)
    \ge \mathbb{F}_C\!\left(\theta \mid M, \rho^{B}_{a|x}(\theta)\right),
\end{equation}
where
\begin{equation}
    \mathbb{F}_C\!\left(\theta \mid M, \rho^{B}_{a|x}(\theta)\right)
    := \sum_{b}
    \frac{\left[\partial_{\theta} p(b|M,\rho^{B}_{a|x}(\theta))\right]^2}
    {p(b|M,\rho^{B}_{a|x}(\theta))}.
    \label{eq:cfi}
\end{equation}

The optimal variance of $H$ is defined as
\begin{equation}
    \Delta H^{opt}:=\min_{x} \sum_{a} P(a|x) \,\Delta[\rho^{B}_{a|x}(\theta),H],
\end{equation} with $\Delta[\rho^{B}_{a|x}(\theta),H]= \rm Tr[H^2~\rho^{B}_{a|x}(\theta)]-Tr[H~\rho^{B}_{a|x}(\theta)]^2$. The condition for Bob's assemblage to be steerable is that it cannot be described by a LHS form as in Eq.~(\ref{eq:LHS}). This leads to the MSI \cite{yadin2021metrological} as $\mathbbm{F}^{opt}_Q \leq 4\,\Delta H^{opt}$ and the metrological steering criterion is
\begin{equation}
    V :=\mathbbm{F}^{opt}_Q - 4\,\Delta H^{opt}.
    \label{eq:MSI}
\end{equation}
Here, $V \le 0$ (i.e., $\mathbbm{F}^{opt}_Q \le 4\,\Delta H^{opt}$) implies that no MSI violation is observed (although this does not by itself imply that the assemblages are not steerable). Conversely, $V > 0$ (i.e., $\mathbbm{F}^{opt}_Q > 4\,\Delta H^{opt}$) certifies that Bob's assemblage is steerable and witnesses a violation of the MSI, with $V > 0$ quantifying the extent of the observed MSI violation.

Note that the violation originates in the incompatibility of Alice's settings, since $\mathbbm{F}^{opt}_Q$ and $\Delta H^{opt}$ are optimized over different $x$. For a single setting the two coincide for pure states and $V=0$. In practice, $\mathbbm{F}_Q$ is replaced by the classical Fisher information $\mathbbm{F}_C$ obtained with a fixed measurement $M$. Since $\mathbbm{F}_C \le \mathbbm{F}_Q$, this substitution can only underestimate $V$, so any violation obtained in this way remains valid.

\subsection{Hidden metrological steering}
The framework developed above assumes ideal, noiseless conditions. We now ask what happens when the phase-encoding process is subject to noise: can the metrological advantage certified by the MSI be recovered? We say that a noisy channel exhibits \textit{hidden metrological steerability} if its assemblage satisfies $V \leq 0$ in the absence of filtering (yielding no certified metrological advantage) yet gives $V > 0$ upon application of suitable stochastic operations.

The unitary operation $U$ in the ideal noiseless scenario is shown in Fig.~\ref{fig:mainfig}(a). However, noise is unavoidable in any quantum communication network, and accounting for it is crucial for obtaining the metrological steering criterion in realistic scenarios (see Fig.~\ref{fig:mainfig}(b)). Hence, we model the noisy phase-encoding operation as
\begin{equation}
    \Lambda_{\theta}(\rho^B_{a|x}) =  \Lambda (U(\theta) \rho^B_{a|x} U^{\dagger}(\theta)).
\end{equation} 
Let us now consider the noise channel $\Lambda$ to be an amplitude damping channel (ADC),
\begin{equation}
    \Lambda(\rho)= \sum_i K_i \rho K^{\dagger}_i,
\end{equation} such that $\sum_i K^{\dagger}_iK_i=\mathbbm{1}$, $K_0 =\begin{pmatrix}
    1 & 0 \\
    0 & \sqrt{(1-\gamma)}
\end{pmatrix}$, $K_1 =\begin{pmatrix}
    0 & \sqrt{\gamma} \\
    0 & 0
\end{pmatrix}$ and $0\le \gamma \le 1$, where $\gamma$ is the amplitude damping parameter. The action of a noisy channel on the phase-encoded quantum state degrades steering, reducing the observed MSI violation. Phase sensitivity also decreases, as reflected in the reduced Fisher information.

We adopt the ADC because it describes the dissipation of energy from a two-level system into its environment, and is therefore the canonical model for decoherence processes in which an excited state decays towards the ground state \cite{nielsen2010quantum}. Its relevance is not restricted to a particular physical implementation. Spontaneous emission in atomic and trapped-ion systems, energy relaxation in solid-state and superconducting qubits, and the decay of an optical mode into a zero-temperature reservoir are all described by the same Kraus decomposition, with $\gamma$ playing the role of the accumulated decay probability along the channel. This is precisely the type of noise that degrades the probe during the phase-estimation process in the metrological platforms mentioned above, so the noise threshold obtained in the following quantifies how much energy relaxation a steering-assisted metrological protocol can tolerate before the certified advantage is lost.

For our considerations in the following, it is important that the unknown phase is generated in the same energy basis that defines amplitude damping; $\Lambda(U\rho U^{\dagger}) = U \Lambda(\rho) U^{\dagger}$ (phase covariance is thus the statement that the channel commutes with the encoding unitary, so that applying the noise before or after the phase encoding yields the same output state). At the level of the Kraus operators, this amounts to $K_i U = e^{i\varphi_i} U K_i$, and the scalar phases $e^{i\varphi_i}$ cancel in $K_i \rho K_i^{\dagger}$, so the condition holds. This phase-covariant alignment allows us to retain the standard generator-based MSI (see Eq.~\ref{eq:MSI}) and isolate the effect of stochastic filtering on the degradation of metrological steering. Crucially, this alignment is not imposed solely for mathematical convenience: it represents the inherent scenario for optical phase estimation under an ADC, and it guarantees that the noisy probe preserves phase information via the same generator.

Phase covariance has an additional implication that will be important when we assess the stochastic filters. Given that the channel commutes with $U(\theta)$, the probability that a filtered event is transmitted remains consistent across all estimated phase values. Consequently, the filters therefore cannot transmit preferentially at specific phases and not others; this mechanism is how parameter-dependent post-selection could shape the measured distribution $p(b|\theta)$ and produce an apparent phase sensitivity. Any observed sensitivity subsequent to filtering is inherently contained within the conditional state itself.

We now show that applying pre- and post-stochastic operations reduces this noise degradation, preserving the channel's quantumness and restoring the violation of the MSI. The pre-stochastic operation $F_{pre}$, applied by Bob before transmission across the ADC, is given by
\begin{equation}
    F_{pre} =  \begin{pmatrix} 1 & 0 \\ 0 & \sqrt{1-f}\end{pmatrix}.
\end{equation} The second operation is performed after the channel action, prior to the Fisher-information-optimizing measurements, i.e., 
\begin{equation}
    F_{post} =  \begin{pmatrix} \sqrt{1-f} & 0 \\ 0 & 1
    \end{pmatrix},
\end{equation} 
where $f$ is the factor representing the strength of the stochastic operation such that $0 \le f \le 1$. Note that both $F_{pre}$ and $F_{post}$ share the same parameter $f$.

Therefore, a noise-robust steering-assisted metrology protocol can now be formulated as follows: 
\begin{enumerate} 
\item Alice performs different measurements and classically communicates the measurement setting and outcome $(x, a)$ to Bob, thereby remotely preparing the corresponding assemblages $\{\rho^B_{a|x}\}$. Bob then applies an encoding unitary $U(\theta)$ before transmission across a phase-covariant ADC $\Lambda$. 
\begin{equation}
\rho_{a|x}^B \xrightarrow {U(\theta)} \rho^B_{a|x}(\theta) = U(\theta) ~\rho^B_{a|x}~U^{\dagger}(\theta).
\end{equation}
Here, the generator of the phase, the Hamiltonian, is chosen as $H = \sigma_z $.

\item Bob then applies the filtering operation $F_{pre}$ before transmission across the ADC $\Lambda$.
\begin{equation}
\begin{split}
\rho^B_{a|x}(\theta) &\xrightarrow {F_{pre}} \rho^{B, f}_{a|x}(\theta) \\
&= \frac{F_{pre}~\rho^B_{a|x}(\theta) ~F_{pre}^\dagger}{\mathrm{Tr}(F_{pre}~\rho^B_{a|x}(\theta)~ F_{pre}^\dagger)}.
\end{split}
\end{equation}

\item The state $\Lambda(\rho^{B, f}_{a|x}(\theta))$ then suffers the amplitude damping effect after which Bob applies another filtering operation $F_{post}$
on this state to obtain
\begin{equation}
\begin{split}
\Lambda(\rho^{B, f}_{a|x}(\theta)) &\xrightarrow {F_{post}} (\rho^{B, f , \gamma}_{a|x}(\theta))' \\
&= \frac{F_{post}~(\rho^{B, f, \gamma}_{a|x}(\theta)) ~F_{post}^\dagger}{\mathrm{Tr}(F_{post}~(\rho^{B, f, \gamma}_{a|x}(\theta))~ F_{post}^\dagger)}.
\end{split}
\end{equation}

\item Bob performs the FI optimizing measurement $M = \{M_b\}$ on the received state $(\rho^{B, f, \gamma}_{a|x}(\theta))'$.
\end{enumerate}

Therefore, the probability distribution of the measurement outcome $b$ at Bob's end is given by 
\begin{equation}
\label{prob}
p(b|M; \theta, f,\gamma) = \mathrm{Tr}\left[M_b~ \left(\frac{F_{post} \circ \Lambda \circ F_{pre} (\rho_{a|x}^{B}(\theta))}{N(\theta,\rho_{a|x}^{B})}\right)\right].
\end{equation}
Here,

\begin{equation}\label{eq:norm}
N(\theta,\rho_{a|x}^{B}) = \mathrm{Tr}\!\left[F_{\mathrm{post}}\,\Lambda_\gamma
\!\left(F_{\mathrm{pre}}U(\theta)\rho_{a|x}^{B} U^\dagger(\theta)F_{\mathrm{pre}}^\dagger
\right) F_{\mathrm{post}}^\dagger \right],
\end{equation}
with $\Lambda_\gamma(\cdot) = \sum_i K_i(\gamma) (\cdot) K_i^\dagger(\gamma)$, 
is the normalization. Since there is alignment between the generator of the phase and the amplitude damping basis, the normalization is independent of the phase $\theta$, i.e., $N(\theta, \rho_{a|x}^{B}) = N(\rho_{a|x}^{B})$.

The normalization $N(\rho_{a|x}^{B})$ is the success probability of the filtering protocol, that is, the fraction of events transmitted by both filters and reaching Bob's measurement stage. As a stochastic operation, the filtering scheme is therefore probabilistic, and the recovery of the MSI violation is obtained at the cost of discarding a fraction $1-N(\rho_{a|x}^{B})$ of the events. This cost is intrinsic to filtering-based activation and has been noted whenever hidden correlations are extracted by local operations \cite{Kwiat2001experimental, ducuara2022hidden}. We emphasize that the question addressed here is not whether steering can be certified from a fair sample, but whether, given that steering survives in the filtered ensemble, it can still be converted into a metrological advantage. The protocol therefore operates on the conditional ensemble by construction, and Eq.~(\ref{eq:norm}) makes the associated cost explicit rather than implicit. Two features keep this post-selection benign for the present purpose. First, as stated above, $N$ is independent of the encoded phase $\theta$, so the filters cannot manufacture phase sensitivity that the state did not already carry. Second, Bob applies the same map to every element of the assemblage, so the filtered assemblage remains non-signaling and the local-hidden-state bound continues to apply to it, as we verify explicitly below. Note that $N$ depends on the conditional state $\rho_{a|x}^{B}$, and hence on Alice's measurement setting, because the states prepared by the $\sigma_z$ setting are aligned with the amplitude damping basis while those prepared by the $\sigma_x$ setting are not. A setting-independent success probability would instead be obtained for the choice $\{\sigma_x, \sigma_y\}$, whose eigenstates lie symmetrically with respect to the damping basis. We keep $\{\sigma_z, \sigma_x\}$ since these settings are the ones that maximize the MSI violation in the noiseless case.

Evaluating the normalization for the protocol gives
\begin{equation}
N(\rho) = (1-f)\left[\rho_{00} + (1-\gamma f)\,\rho_{11}\right],
\label{eq:success}
\end{equation}
where $\rho_{00}$ and $\rho_{11}$ are the populations in the amplitude damping basis. Since Bob applies the same map to every element of the assemblage, $\sum_a N(\rho^B_{a|x}) = (1-f)(1-\gamma f/2)$ for both settings, so the filtered assemblage remains non-signaling. The individual weights, however, are no longer uniform for $x=0$, where they become $1/(2-\gamma f)$ and $(1-\gamma f)/(2-\gamma f)$. Throughout this work we evaluate $\mathbbm{F}^{opt}_Q$ and $\Delta H^{opt}$ with the original weights $P(a|x)=1/2$. This choice leaves $\mathbbm{F}^{opt}_Q$ unchanged, because the $x=0$ states commute with $H=\sigma_z$ and contribute no Fisher information, so the maximum over $x$ is always attained at $x=1$, where the two conditional states have equal success probability and the weights remain uniform. It does affect $\Delta H^{opt}$, which is attained at $x=0$, and since $(1-\gamma f)/(2-\gamma f) \le 1/2$, using uniform weights can only overestimate $\Delta H^{opt}$ and therefore underestimate $V$. The violations and the critical damping parameters reported below are thus conservative, and constitute lower bounds on the values obtained with the updated weights.

Fig.~\ref{fig:mainfig}(c) depicts the schematic of the protocol. Here, we take Alice's measurements to be \{$\sigma_z$, $\sigma_x$\}, which yield the four eigenstates of $\sigma_z$ and $\sigma_x$. We apply the phase-shift Hamiltonian $H=\sigma_z$ to these four states. Bob's measurement is set to $M=\sigma_y$. With this specification, applying the noise channel and the filters, we calculate the classical Fisher information (see Eq.~\ref{eq:cfi}) and the optimal variance of the Hamiltonian. We find that the MSI violation can be restored for increasing values of $f$, as we confirm experimentally in the following section.


\subsection{Experimental activation of hidden metrological steerability}
We tested the predictions above in a photonic implementation, sketched in Fig.~\ref{fig:setup} and described in detail in the Methods section. Alice's four conditional states are encoded in the polarization of heralded single photons, the phase is imprinted with a liquid-crystal variable retarder, the ADC is realized in a two-path Sagnac interferometer, and the stochastic filters $F_{pre}$ and $F_{post}$ are stacks of thin glass plates at the Brewster angle.

To experimentally demonstrate the activation of metrological steering by applying the proposed stochastic filters, we measured the MSI violation parameter $V$ of Eq.~(\ref{eq:MSI}) as a function of $\gamma$ for a fixed phase $\theta = \pi$. Different values of $\gamma$ are set by the half-wave plate angle $\theta_1$ inside the Sagnac interferometer (see Methods). The conditional probabilities $p(b|M,\rho^B_{a|x}(\theta))$ are obtained from the coincidence counts recorded between the trigger detector and Bob's two detectors.

\begin{figure}[t!]
\centering
\includegraphics[width=0.65\textwidth]{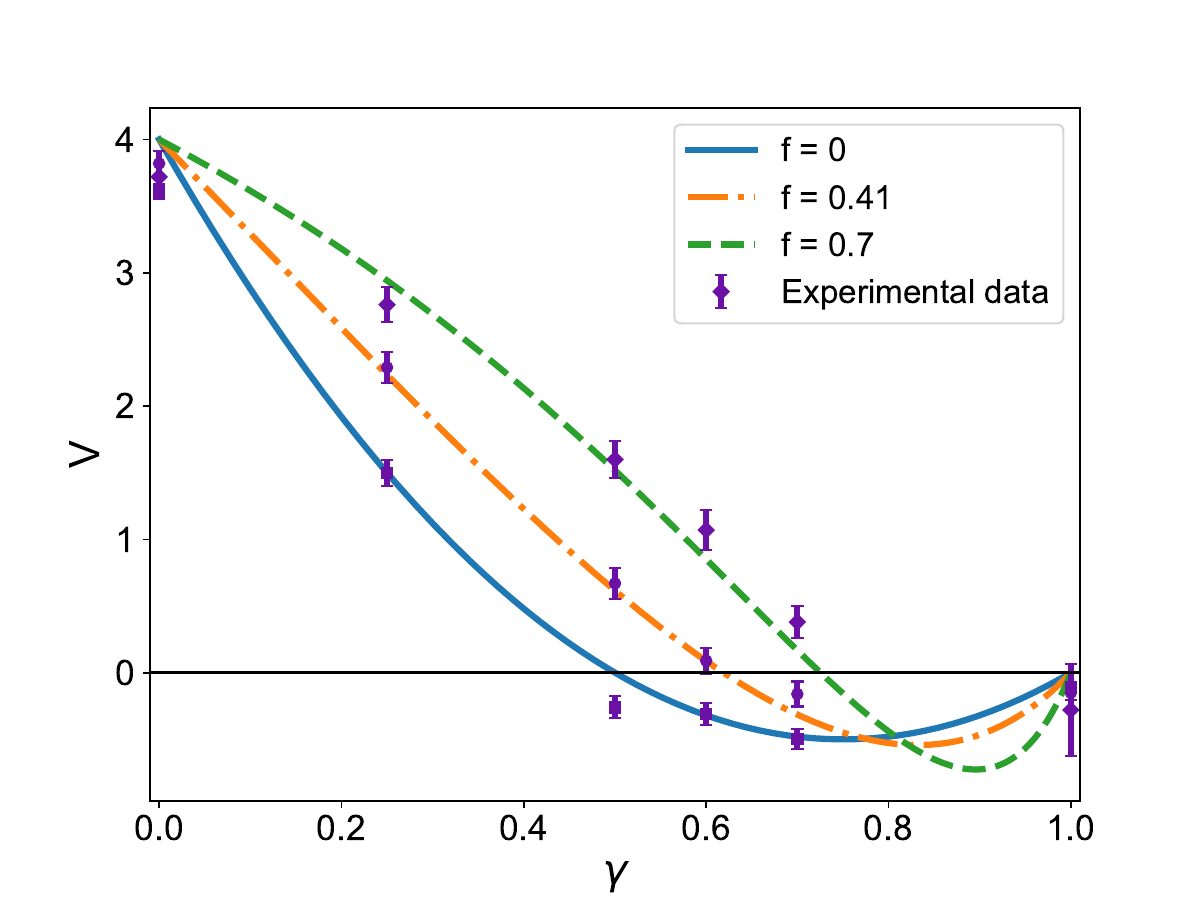}
\caption{The MSI violation parameter $V$ as a function of the amplitude-damping parameter $\gamma$ for a fixed phase $\theta=\pi$. The solid (blue), dot-dashed (orange), and dashed (green) lines represent the theoretical predictions when $f = 0$, $0.41$, and $0.70$, respectively. The purple data points are the experimental results for the three filter settings, each following its corresponding theoretical curve. The error bars are obtained using Gaussian error propagation and Poisson statistics for the coincidence counts. Observe that, depending on the filter parameter values, an increase in the corresponding $\gamma_c$ can be achieved, indicating that the protocol becomes more robust against ADC when stochastic filters are applied.}
\label{fig:results}
\end{figure}

Fig.~\ref{fig:results} shows $V$ as a function of $\gamma$. The solid, dot-dashed, and dashed lines correspond to the theoretical predictions when $f = 0$, $0.41$, and $0.70$, respectively. The purple data points are the experimental results, measured for the three filter settings, each set following its corresponding theoretical curve. The error bars are obtained using Gaussian error propagation and considering the Poisson statistics of the recorded coincidence counts.

In the case without filtering ($f = 0$), $V$ decreases rapidly with $\gamma$, falling below zero at $\gamma_c \approx 0.5$. In contrast, with $f = 0.41$ ($f = 0.70$), $V$ can beat the LHS bound until $\gamma_c \approx 0.6$ ($\gamma_c \approx 0.75$), beyond which the scheme no longer shows improvement. The experimental data agree well with the theoretical predictions. Small deviations between the experimental points and the theoretical curves can be attributed to inaccuracies in the wave-plate angles used to implement the ADC and the measurements, as well as to the finite-difference approximation used to estimate the derivative of the probability distribution. The experimental data also decrease rapidly as a function of $\gamma$. However, when $F_{pre}$ and $F_{post}$ are applied, the noise tolerance is increased, extending the range of $\gamma$ for which the MSI violation can be maintained. These results experimentally confirm the activation of hidden metrological steerability: channels that yield no certified metrological advantage without filtering recover a certified steering advantage through the application of stochastic operations.

\section{Discussion}
We have introduced a theoretical framework for the recovery of metrological advantage in steering scenarios, showing that stochastic pre- and post-filtering operations activate a metrological steering advantage that is provably absent in the unfiltered channel, and we identified the parameter regime in which this hidden advantage becomes accessible. The filters act directly on the quantity that certifies the metrological task, the violation of the MSI. We modeled the phase-encoding channel as an ADC because it describes energy dissipation from a two-level system into its environment, the dominant decoherence mechanism in the platforms where steering-assisted metrology would be deployed. Spontaneous emission in atomic and trapped-ion sensors, energy relaxation ($T_1$) in solid-state qubits, and the decay of an optical mode into a zero-temperature reservoir are all described by the same map, so the noise threshold we report quantifies how much energy relaxation a steering-based metrological protocol can tolerate before stochastic filtering becomes necessary.

We confirmed the practicality of this framework experimentally, using polarization-encoded photonic qubits transmitted through an ADC implemented with a two-path Sagnac interferometer. Without filtering, the MSI violation vanishes at $\gamma_c \approx 0.5$. With $f = 0.41 \pm 0.05$ and $f = 0.70 \pm 0.02$, $\gamma_c$ rises to $\approx 0.6$ and $\approx 0.75$, respectively, in agreement with the theoretical predictions. The recovery mechanism is not tied to amplitude damping, and the conditions a noisy channel must satisfy for it to apply can be stated explicitly. Two ingredients are required. The channel must be covariant with respect to the generator of the encoded phase, so that the generator-based MSI is retained and the success probability of the filters is independent of the parameter being estimated. The channel must also possess a Kraus structure in which the population transfer occurs between the same basis states that the filters attenuate, so that a pre-filter suppressing the state most vulnerable to the noise can be undone by a post-filter acting on its complement. Amplitude damping satisfies both conditions in the simplest possible way, which is why we adopt it here, but any phase-covariant channel with a preferred decay direction, such as generalized amplitude damping at finite temperature \cite{PhysRevA.95.042342}, meets them as well. Channels lacking a preferred direction, such as depolarizing noise, would require a different filter construction and fall outside the present treatment. Finally, given the interest in steering as a resource in quantum information protocols, our recovery framework is directly relevant to other applications, such as one-sided device-independent quantum key distribution or noise adaptation for the certification of quantum channels. Our results motivate further experimental investigation of noise-adapted metrological protocols for other canonical noise channels.

\section{Methods}
\subsection{Photonic implementation}
\begin{figure}[t!]
\centering
\includegraphics[width=0.8\textwidth]{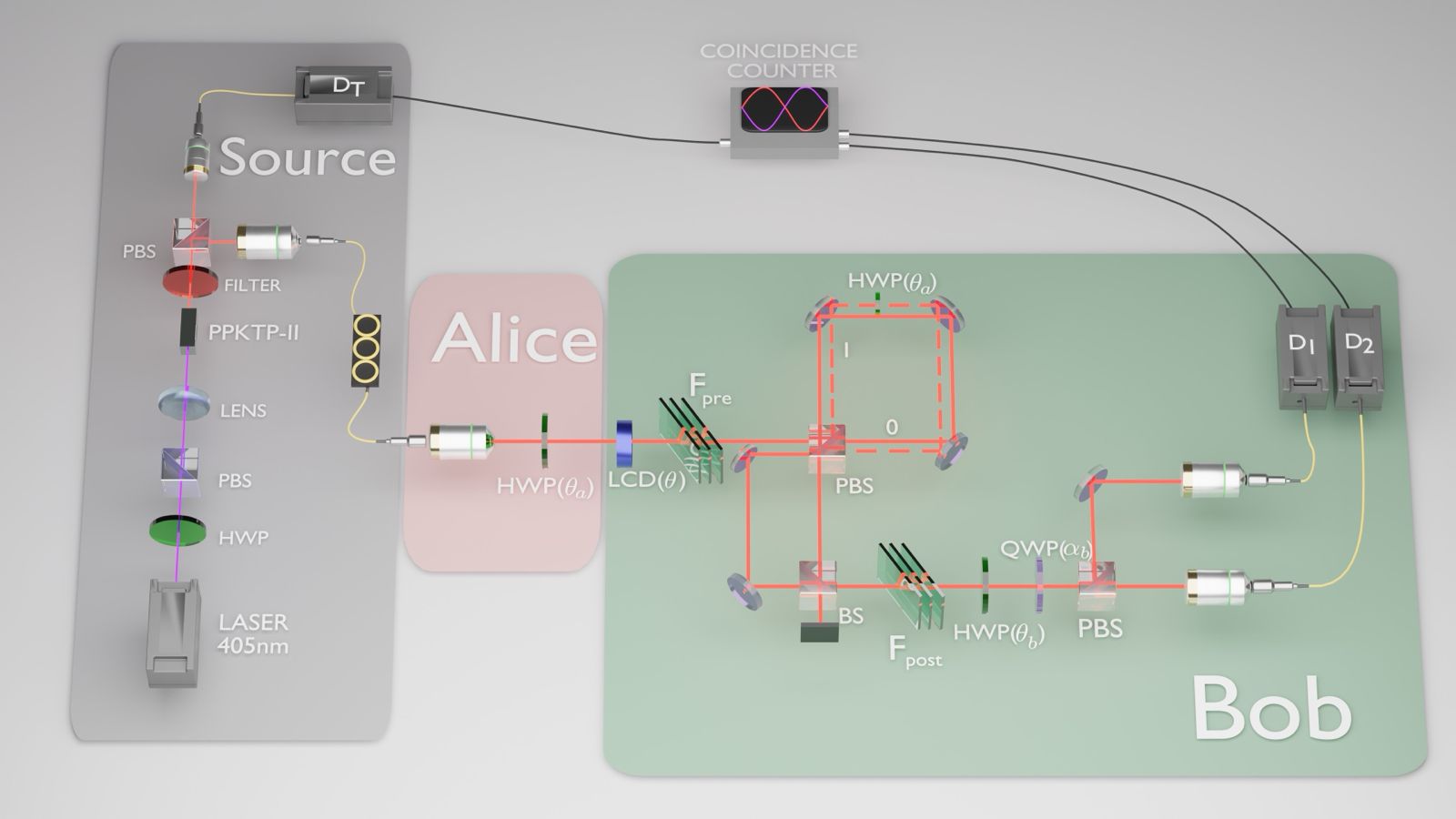}
\caption{Photonic implementation used to study the behavior of the metrological steering protocol in a noisy scenario. A heralded photon source based on the SPDC process is used to produce single photons. Alice's setup includes an HWP($\theta_a$) that is used to prepare four states encoded in the polarization degree of freedom: $\rho_{0|0}$, $\rho_{0|1}$, $\rho_{1|0}$, and $\rho_{1|1}$. The unitary phase encoding is then applied via the LC variable retarder LCD ($\theta$), followed by the pre-filter $F_{pre}$. Subsequently, a two-path Sagnac interferometer implements the ADC, where $\gamma$ is controlled by $\theta_1$. After transmitting the state through the channel, Bob applies $F_{post}$ and, finally, two measurements are performed by a QWP, HWP, a PBS, and two avalanche photon detectors. The stochastic filters $F_{pre}$ and $F_{post}$ are implemented using thin glass plates placed at the Brewster angle.}
\label{fig:setup}
\end{figure}

Our noise-adapted demonstration of the metrological steering protocol is depicted in Fig.~\ref{fig:setup}. The experiment involves: (i) state preparation, in which Alice encodes four states onto the polarization of single photons; (ii) a unitary phase encoding $U(\theta)=e^{-i\sigma_z\theta}$ applied by Bob; (iii) a noisy channel (ADC) through which the polarization qubits travel; (iv) pre- and post-filtering operations $F_{pre}$ and $F_{post}$, applied by Bob before and after the ADC to mitigate the effects of noise; and (v) projective measurements by Bob to test the MSI.

Single photons are produced using a heralded single-photon source based on the spontaneous parametric down-conversion (SPDC) process~\cite{karthik2025nrqrac}. During the SPDC process, a pair of photons with orthogonal polarizations at 810~nm is generated. A polarizing beam splitter (PBS) separates the photons based on their polarization. One of the down-converted photons arrives at an avalanche photodetector $D_T$, which acts as a trigger, while the other photon is sent to Alice's section. Single-mode fibers (SMFs) are used to couple the down-converted photons to the detectors and also project onto a single spatial mode, thus removing residual spatial correlations. To maintain the polarization of the photons during propagation through the SMFs, a manual fiber polarization controller (MPC) is employed. A half-wave plate (HWP) at angle $\theta_a$ relative to its fast axis is used to prepare the four polarization qubit states $\rho_{a|x}$
\begin{align}
\rho_{0|0} &= \ket{H}\bra{H}, & \rho_{1|0} &= \ket{V}\bra{V}, \nonumber \\
\rho_{0|1} &= \ket{D}\bra{D}, & \rho_{1|1} &= \ket{A}\bra{A},
\label{eq:states}
\end{align}
where $\ket{H}$ and $\ket{V}$ are the horizontal and vertical polarization modes, respectively. Similarly, $\ket{D}=(\ket{H}+\ket{V})/\sqrt{2}$ and $\ket{A}=(\ket{H}-\ket{V})/\sqrt{2}$ correspond to the diagonal and anti-diagonal polarization modes. These are the eigenstates of $\sigma_z$ ($x=0$) and $\sigma_x$ ($x=1$), representing Alice's two incompatible measurement settings in the temporal steering scenario.

The unitary phase encoding $U(\theta) = e^{-i\sigma_z\theta}$ corresponds, up to a global phase, to a phase retarder:
\begin{equation}
e^{-i\sigma_z\theta} = e^{i\theta}\begin{pmatrix} e^{-2i\theta} & 0 \\ 0 & 1 \end{pmatrix}.
\label{eq:retarder}
\end{equation}
The optical retardance is $\varphi = 2\theta$.
This operation is implemented using a liquid crystal (LC) variable retarder (Thorlabs Full-Wave LC Retarder), a birefringent device whose retardance is controlled by the applied AC voltage \cite{saleh2019fundamentals}. By varying the voltage, we control the phase shift $\theta$ imprinted on the horizontal component of the polarization state while leaving the vertical component unchanged. The retarder is calibrated at the operating wavelength of 810~nm by identifying the voltage corresponding to a full-wave ($\lambda$) retardance, $\varphi = 2\pi$, which by Eq.~(\ref{eq:retarder}) corresponds to $\theta = \pi$ up to a global phase. In the optical path, the LC retarder is placed immediately after Alice's state preparation and before the pre-filter $F_{pre}$, so that $U(\theta)$ acts on the conditional state $\rho^B_{a|x}$ prior to any filtering or noise (see Fig.~\ref{fig:mainfig}).

The ADC, represented as a decomposition of the Kraus operators $K_0(\gamma)$ and $K_1(\gamma)$, is implemented by a two-path Sagnac interferometer \cite{almeida2007environment,salles2008experimental}, by incoherently combining the two outputs of the interferometer. The photon first interacts with a PBS, which separates the horizontal and vertical components into two linear-momentum modes, $\ket{0}$ and $\ket{1}$, depicted as continuous and dashed lines in Fig.~\ref{fig:setup}, respectively. Inside the interferometer, an HWP($\theta_1$) is added on path $1$ with its fast axis aligned at angle $\theta_1$, while a compensating HWP at $0^\circ$ is placed on path $0$ to equalize the optical path lengths (not shown in the figure). The two paths are routed using three mirrors so that the photon passes through the same PBS again. After the PBS, the spatial mode $\ket{1}$ travels a longer distance than $\ket{0}$, and both modes are recombined at a beam splitter (BS), where the optical path length difference exceeds the photon coherence length. This incoherent recombination implements the partial trace over the environment (linear momentum degree of freedom), yielding the output density matrix
\begin{equation}
\rho_{\mathrm{out}} = K_0\ket{\psi}\bra{\psi}K_0^\dagger + K_1\ket{\psi}\bra{\psi}K_1^\dagger,
\end{equation}
where the Kraus operators are given by
\begin{align}
K_0(\theta_1) &= \ket{H}\bra{H} + \cos(2\theta_1)\ket{V}\bra{V}, \nonumber \\
K_1(\theta_1) &= \sin(2\theta_1)\ket{H}\bra{V}.
\end{align}
Comparing with the standard ADC Kraus operators, we see that $\sin(2\theta_1)=\sqrt{\gamma}$ and $\cos(2\theta_1)=\sqrt{1-\gamma}$, allowing for control of $\gamma$ using the wave-plate angle $\theta_1$. When $\theta_1=0$, all photons leave the Sagnac interferometer through output $\ket{0}$, indicating no amplitude attenuation in the channel, as expected. When $\theta_1 \neq 0$, photons start to populate the $\ket{1}$ mode, indicating an increase in noise. 

The filter operations $F_{pre}$ and $F_{post}$ are implemented using thin glass plates placed at the Brewster angle, so that only the polarization component orthogonal to the plane of incidence is partially reflected. Thus, in the transmitted beam, which is the one used here, one polarization mode is partially attenuated while the other is left unchanged \cite{Kwiat2001experimental}. In the optical path, $F_{pre}$ is placed after the unitary phase encoding and before the ADC, while $F_{post}$ is placed after the ADC and before Bob's measurement stage. The difference between $F_{pre}$ and $F_{post}$ lies in the position of the principal axis of the glass plates, which are orthogonal to each other, allowing the vertical or horizontal components to be filtered. The filter parameter $f$ is controlled by the number of glass plates in the stack. We consider two sets of glass plates: three and eight glass plates, corresponding to filter parameters of $f = 0.41 \pm 0.05$ and $f = 0.70 \pm 0.02$, respectively. More details on the characterization of the filters are given in Ref.~\cite{karthik2025nrqrac}.

Bob's measurements for extracting the MSI violation parameter in our protocol are given by $\sigma_z$ and $\sigma_y$, where $\sigma_z = \ket{H}\bra{H} - \ket{V}\bra{V}$ and $\sigma_y = \ket{L}\bra{L}-\ket{R}\bra{R}$ are the Pauli matrices, with $\ket{R}=(\ket{H}-i\ket{V})/\sqrt{2}$ and $\ket{L}=(\ket{H}+i\ket{V})/\sqrt{2}$. The measurements are implemented using a QWP($\alpha_b$) and HWP($\theta_b$), followed by a PBS and two avalanche single-photon detectors $D_1$ and $D_2$, as shown in Fig.~\ref{fig:setup}. The $\sigma_y$ measurement is used to evaluate the classical Fisher information via Eq.~(\ref{eq:cfi}), while $\sigma_z$ is used to compute the optimal variance of the Hamiltonian $H=\sigma_z$. The derivative of the probability distribution entering the classical Fisher information is computed numerically using the central difference approximation $\partial_\theta p \approx [p(\theta+h)-p(\theta-h)]/(2h)$, with $h=2.5^\circ$ in $\theta$, implemented as controlled deviations $\pm 2h$ in the retardance of the LC retarder. The signals from $D_T$, $D_1$, and $D_2$ are processed by a coincidence counting module with calibrated delays between the outputs of $D_1$ ($D_2$) and $D_T$. To reduce accidental counts, the module is set to a 1~ns coincidence window. Each coincidence count is recorded over a 10~s integration time.


\backmatter

\bmhead{Acknowledgements}
HSK, MP thank NCN Poland, ChistEra-2023/05/Y/ST2/00005 under the project Modern Device Independent Cryptography (MoDIC).
ASH acknowledges full support, while MP acknowledges that this work is partially carried out under IRA Programme, project no. FENG.02.01-IP.050006/23, financed by the FENG program 2021-2027, Priority FENG.02, Measure FENG.02.01., with the support of the FNP.
This research was funded by ANID Anillo Project ATE250003, Fondo Nacional de Desarrollo Cient\'{\i}fico y
Tecnol\'ogico (FONDECYT) Grant No.\ 1260111, 1240746, 1231940, and ANID Millennium Science Initiative Program ICN17--012.

\section*{Declarations}

\bmhead{Funding}
See Acknowledgements.

\bmhead{Competing interests}
The authors declare no competing interests.

\bmhead{Data availability}
The datasets used and/or analyzed during the current study are available from the corresponding author upon reasonable request.

\bmhead{Author contributions}
H.S.K., A.S.H. and M.P. developed the theory. J.B., D.M., J.P.-P. and F.M.Q. performed the experiment and carried out the measurements. S.P.W., G.L. and E.S.G. supervised the experimental setup. J.B. and E.S.G. analyzed the results. H.S.K. and E.S.G. wrote the manuscript. All authors discussed the results and approved the final version of the manuscript.

\bibliography{reference}

\end{document}